\documentclass[10pt, twocolumn]{article}

\usepackage[font=small,labelfont=bf]{caption}
\usepackage{graphicx}
\usepackage{amsmath}
\usepackage{bm}
\usepackage{amsfonts}
\usepackage{amssymb}
\usepackage{mathtools}
\usepackage{mathrsfs}
\usepackage{multirow}
\usepackage{url}
\usepackage{hyperref}
\usepackage{algorithm}
\usepackage{algpseudocode}
\usepackage[draft]{fixme}
\usepackage{relsize}
\usepackage{authblk}

\makeatother

\newcommand{\tv}[1]{\textbf{#1}}

\newcommand*{\figref}[2][]{%
  \hyperref[{#2}]{%
    \ref*{#2}%
    \ifx\\#1\\%
    \else
      #1%
    \fi
  }%
}

\begin{document}

\title{Data-driven model order reduction for granular media}

\author[1]{Erik Wallin}
\author[1]{Martin Servin}
\affil[1]{Ume{\aa} University}

\date{\today}

\maketitle

\begin{abstract}
We investigate the use of reduced-order modelling to run discrete element simulations at higher speeds. Taking a data-driven approach, we run many offline simulations in advance and train a model to predict the velocity field from the mass distribution and system control signals. Rapid model inference of particle velocities replaces the intense process of computing contact forces and velocity updates. In coupled DEM and multibody system simulation the predictor model can be trained to output the interfacial reaction forces as well. An adaptive model order reduction technique is investigated, decomposing the media in domains of solid, liquid, and gaseous state. The model reduction is applied to solid and liquid domains where the particle motion is strongly correlated with the mean flow, while resolved DEM is used for gaseous domains. Using a ridge regression predictor, the performance is tested on simulations of a pile discharge and bulldozing. The measured accuracy is about 90\% and 65\%, respectively, and the speed-up range between 10 and 60.
\end{abstract}

\section{Introduction}
\label{sec:introduction}
Computational modelling of granular dynamics has important applications in both science and engineering, but is challenging due to the complex nature of granular media.
The discrete (or distinct) element method (DEM) is perhaps the most versatile numerical method for it.
It supports the three granular phases, solid, liquid, and gas.
It can capture both discrete and collective phenomena, that depend on contact parameters, particle shape and arrangements.
DEM simulations are, however, computationally intense, which limits the practical applicability.

There are basically three methods of accelerating a DEM simulation.
Firstly, the computational speed can be increased by parallelization and use of specialized hardware \cite{Preclik2015,Steuben2016,Gan2016,Tian2017,He2018,Rakotonirina2018}, but the monetary cost and energy consumption grows rapidly with system size.
Secondly, changing from explicit to implicit time-integration allows for much larger time-steps than the limit set by the time-period of free vibrations for particles of given mass and contact stiffness.
The computational bottleneck is then shifted from collision detection to solving the equations of motion and contact force computation.
Depending on the system properties and error tolerance this may be very advantageous \cite{servin:2014:esn}.

The third way is to employ some form of model order reduction, where the original system is substituted with an approximation that require fewer variables and computational operations per simulated unit of time.
Normally, model order reduction is seen as a projection from a high-dimensional space to a low-dimensional subspace, where the time-integration can be performed with manageable computational intensity.
Once advanced in time, the solution can be projected back to the original high-dimensional space.
The process introduces a model reduction error, that may or may not be acceptable for the intended purpose of the simulation.

In the present paper we explore the possibility of accelerating DEM simulation using data-driven model reduction.
The idea is to perform numerous detailed simulations of a system in advance, train a model to predict new system states and use these to advance a running simulation faster in time than the original simulations.
The question is whether and how this is at all feasible, and what speed-up and accuracy can be achieved.

The data-driven approach has the disadvantage that a certain amount of resolved simulations must be performed in advance to generate training data for building a model. 
In fact, a new model must be generated for each confining geometry and set of material parameters. 
The question is whether the advantages of the method can outweigh this drawback.
Simulations that must run at real-time speed is one type of application that may benefit from using this technique.
Specific examples are simulators for operator training, system testing with hardware-in-the-loop or embedded simulations serving a model-based controller.
Another class of applications is surrogate models \cite{Forrester2008} for simulation-based planning and optimization in parameter spaces too large to be covered with full-resolution simulation, but manageable with a reduced-order model trained on a comparatively small set of resolved simulations.

To increase the knowledge about the challenges and opportunities of accelerating DEM simulations using data-driven model order reduction, we have developed and tested a realisation of this idea.
First, a general method is described, and error measures are introduced. 
Next, this is implemented, using a ridge regression model for predicting the velocity field.
The computationally expensive process of computing contact forces is substituted by rapid inference of the model to output the velocity field at the particle positions.
This approach can be expected to perform well in the solid and liquid regime but poorly in the gaseous regime, where individual particle motion is not strongly correlated with the mean flow.
Therefore, we investigate an adaptive model order reduction technique, where the system is decomposed in solid/liquid, and gaseous parts.
Gaseous sub-domains are integrated using full resolution DEM while the reduced-order model is applied to the solid/liquid sub-domains.
The method is tested on two different systems, a pile with controlled feed and gravity-driven discharge flow, and a blade cutting and pushing through a particle bed like a bulldozer blade.
In both cases a model is trained to predict the velocity field from the given input signals and the current mass distribution.
In the bulldozer case, the force on the blade is also predicted.
The accuracy and computational speed-up are analysed on these systems.

We are motivated by, but do not explore, the opportunities with deep learning, that shows promising results for predicting the velocity field in fluid dynamics \cite{Kutz2017,Brunton2020}. The present work is a first step in that direction for granular media. As such, it is natural to investigate the performance of a plain regression model and building knowledge for employing more advanced machine learning algorithms.

\subsection{Previous work}
In the DEM literature there are only a few examples of model order reduction.
Boukouvala et al. \cite{Boukouvala2013} explored discrete element reduced-order modelling for particle mixing in a blade blender.
By principal component analysis (PCA) of simulation snapshots, sampled in a regular grid covering the mixer interior, models were built for predicting the particle velocity field and blade force as function of the mixer control parameters (blade speed and geometry).
In turn this was used to develop a surrogate model for optimization of mixing performance based on a relatively small set of time-consuming DEM simulations.
This work was later extended by Rogers et al. \cite{Rogers2014}, considering also the effect of dynamic response from changes in the control parameters.
The reduced-order model was not used to accelerate the DEM simulations themselves.

In \cite{servin:2016:amr}, Servin and Wang developed an adaptive model order reduction technique which substitute particles that collectively move as a single rigid body, with six degrees of freedom rigid aggregates of the corresponding mass, momentum, and contact shape.
Different strategies for predicting when the aggregate should split in smaller constituents was investigated.
The method is severely limited by that the reduced model support only rigid body modes of motion.

Recently, Zhong and Sun \cite{Zhong2018}, with inspiration from \cite{Williams1997} and \cite{Lee2013}, investigated a reduced-order model for granular materials under small viscoelastic deformations, with fix connectivity between the particles, using proper orthogonal decomposition (POD) of the displacement field. 

Pseudo-particle modelling \cite{Feng:DEM:2014} may be considered a form of model order reduction.
Each pseudo-particle represents the collective effect of many small real particles.
The pseudo-particle shape and contact parameters are considered model parameters that are calibrated to give the material the approximate bulk-mechanical properties of the original system.
We have not found any example in the DEM literature where the state of a fine-resolution particle model is projected to a pseudo-particle subspace and projected back after time-integration.
Ideas of this type may be found in the literature of computer graphics \cite{Ihmsen2013} for the purpose of visual appearance and with no analysis of the reduction error.

\section{Model order reduction}
Let us first summarize the classical meaning of model order reduction \cite{Antoulas2005}.
Consider a dynamical system
\begin{align}
  \dot{\bm{x}} & = \bm{f}(\bm{x},\bm{u}) \\
  \bm{y} & = \bm{g} (\bm{x},\bm{u})
\end{align}
with vectors of state $\bm{x}(t) \in \mathbb{R}^n$, input $\bm{u}(t) \in \mathbb{R}^m$ and output $\bm{y}(t) \in \mathbb{R}^q$.
The problem of model order reduction can be stated as follows.
Find a lower order model
\begin{align}
  \dot{\hat{\bm{x}}} & = \hat{\bm{f}}(\hat{\bm{x}},\bm{u}) \\
  \hat{\bm{y}} & = \hat{\bm{g}} (\hat{\bm{x}},\bm{u})
\end{align}
that produce approximately the same observations
\begin{equation}
  \left\Vert \bm{y} - \hat{\bm{y}} \right\Vert \leq \varepsilon_\text{r} \left\Vert \bm{u} \right\Vert
\end{equation}
to an accuracy $\varepsilon_\text{r}$ for all input signals $\bm{u} \in \mathcal{U}$ relevant to the particular application.
The idea is that the subspace where the reduced state vector lives, $\hat{\bm{x}}(t) \in \mathbb{R}^r$, is of much lower dimension than the original system space, i.e., $r \ll n$.  Note that the observation vector, $\hat{\bm{y}} \in \mathbb{R}^q$, must have the same dimensions as in the original system or there must exist a projection operator to that space.

The standard methods for computing approximate low-order models are the SVD-based and Krylov-based approximation methods \cite{Antoulas2005}.
The proper orthogonal decomposition (POD) method is a special case of SVD-based model order reduction that is particularly popular in computational mechanics and fluid dynamics.
However, granular media modelled using DEM differ from many other dynamical and physical systems in that the connectivity of the variables changes frequently and unpredictably.
Therefore, a non-standard model order reduction approach is necessary.

\section{A reduced-order discrete element method}
\label{sec:RODEM}

In this section we describe a reduced-order model for granular media simulation using the discrete element method.

\subsection{Resolved DEM}
\label{sec:resolvedDEM}
We first briefly describe the standard discrete element method. We will refer to this as \emph{resolved DEM}.

Each of the $N_\text{p}$ particles, indexed $a \in \mathcal{N}$, has a position $\bm{x}^a(t) \in \mathbb{R}^3$, velocity $\bm{v}^a(t) = \dot{\bm{x}}^a$, scalar mass $m^a$ and diameter $d^a$.
For clarity of the exposition, we ignore the rotational degrees of freedom.
The equations of motion are
\begin{align}
	\dot{\bm{x}} & = \bm{v} 						\label{eq:x_dot}\\
	\bm{M} \dot{\bm{v}} & = \bm{f}(\bm{x},\bm{v},t)	\label{eq:v_dot}
\end{align}
with system position vector $\bm{x} \in \mathbb{R}^{3N_\text{p}}$, velocity vector $\bm{v} \in \mathbb{R}^{3N_\text{p}}$, diagonal mass matrix $\bm{M} \in \mathbb{R}^{3N_\text{p} \times 3N_\text{p}}$.
The force $\bm{f}$ is the sum of external forces and contact forces.
Each of the $N_\text{c}$ particle-particle contacts, indexed by $n \in \mathcal{N}_\text{c}$, have a contact position $\bm{x}_{\text{c},n}$ and pairwise contact force $\bm{f}_n^{ab} \in \mathbb{R}^3$ on particle $a$ from particle $b$.
Each contact force has one normal and two tangential (friction) components.
The computationally intense part is the numerical integration of the velocity which involve contact detection and contact force computation.
Thus, we ascribe the system a \emph{computational dimensionality} of $N_\text{d} = 3N_\text{p} + 3N_\text{c}$. This is simply the number of equations of motion and contact force equations to be solved during each simulation time-step. 
If rotational degrees of freedom are included, as well as rolling resistance, this changes to $N_\text{d} = 6N_\text{p} + 6N_\text{c}$.
For rough estimates it can be assumed that each particle have up to 10 contacts with neighbouring particles, i.e $N_\text{c} \lesssim 5 N_\text{p}$.

In DEM, numerical integration is normally performed using an explicit method with small time-steps, that resolve the natural oscillation frequency given the particle mass and stiffness.
The computational bottleneck is the collision detection and force calculation from the contact overlap and relative velocity.
For some systems it is more beneficial to use the time-implicit method referred to as nonsmooth contact dynamics, or nonsmooth DEM \cite{Radjai:2009:cdn,servin:2014:esn}.
This allows for large time-step integration, moving the computational bottleneck to the process of solving the nonlinear equations (variational inequalities) from the Signorini-Coulomb and Newton contact laws.
The computational complexity and speed, stands in direct relation to the computational dimensionality, but the explicit and implicit methods scales differently \cite{servin:2014:esn}.

\subsection{Reduced DEM}
\label{sec:reducedDEM}
Let the particles be divided in two subsystems, $A$ and $B$, such that $\bm{x} = [\bm{x}_A, \bm{x}_B]$ and $\bm{v} = [\bm{v}_A, \bm{v}_B]$.
The force is divided as $\bm{f} = [\bm{f}_A + \bm{f}_{AB}, \bm{f}_B + \bm{f}_{BA}]$, where $\bm{f}_{AB}$ is the interfacial forces on $A$ from $B$, and $\bm{f}_A$ and $\bm{f}_B$ denote forces acting only on particles within $A$ and $B$, respectively.
The computational dimensionality of the system can be decomposed as $N_\text{d} = 3 ( N^A_\text{p} + N^B_\text{p}) + 3 ( N^A_\text{c} + N^B_\text{c} + N^{AB}_\text{c})$.
Assume that the particles in subsystem $B$ move according to a known velocity field $\bm{u}(\tv{x},t)$.
Each particle $b \in \mathcal{N}_B$ thus has a known velocity $ \bm{v}_b = \bm{u}(\tv{x},t)_{\tv{x} = \bm{x}_b}$ at coordinate $\tv{x}$ and time $t$.
We abbreviate this as $\dot{\bm{x}}_B = \bm{u}(\bm{x}_B,t)$.
Consequently, the particle acceleration is $\dot{\bm{v}}_b = \partial_t \bm{u}(\bm{x}_b,t) + \bm{v}_b \cdot \nabla \bm{u}(\tv{x},t)_{\tv{x} = \bm{x}_b}$.
This eliminates the equation for $\dot{\bm{v}}_B$ and leave us with the following, reduced, set of equations of motion
\begin{align}
	\begin{bmatrix}
		\dot{\bm{x}}_A \\
		\dot{\bm{x}}_B
	\end{bmatrix}
	& =
	\begin{bmatrix}
		\bm{v}_A \\
		\bm{u}(\bm{x}_B,t)
	\end{bmatrix}
	\label{eq:x_evolution} \\
	\bm{M}_A \dot{\bm{v}}_A  & =\bm{f}_A + \bm{f}_{AB} \label{eq:v_A_evolution}.
\end{align}
If the velocity field $\bm{u}(\tv{x},t) $ can be computed with negligible effort, the computational intensity of the system is that of integrating Eq.~(\ref{eq:v_A_evolution}).
The dimensionality of the reduced model is then $N'_\text{d} = 3 N^A_\text{p} + 3 ( N^A_\text{c} + N^{AB}_\text{c})$.
Assuming each particle is in contact with a handful of other particles the \emph{reduction factor} become $R \equiv N'_\text{d} / N_\text{d} \gtrsim N^A_\text{p} / N_\text{p}$. 
In our implementation and tests, we include rotation in resolved DEM but not in reduced DEM.

\begin{figure}
    \centering
    \includegraphics[width=0.4\textwidth]{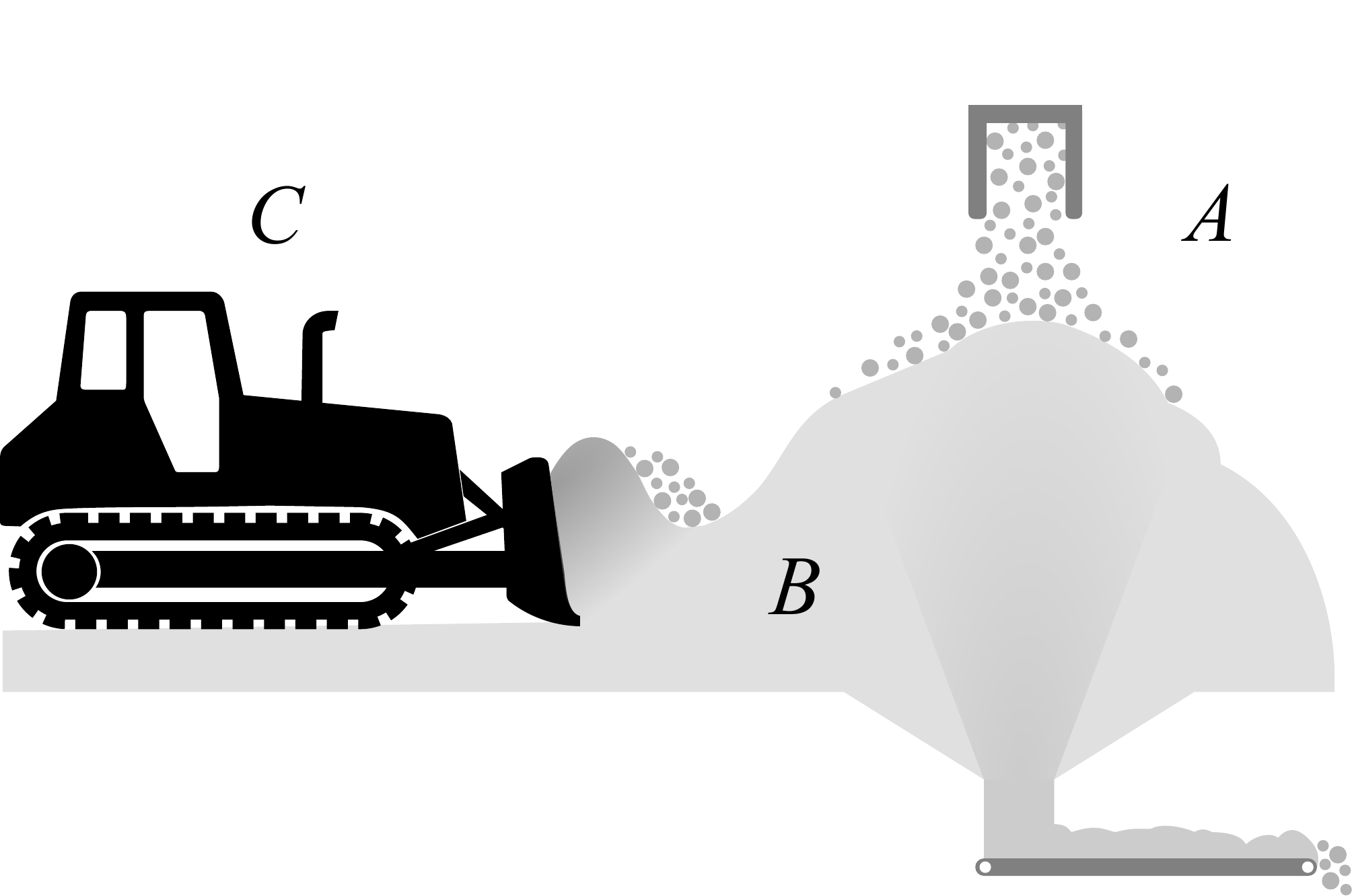}
	\caption{Illustration of a granular system divided in a high-resolution part (A), reduced-order part (B)
	and coupling with a multibody system (C).}
    \label{fig:system_illustration}
\end{figure}

\subsection{Model reduction errors}
There can be two sources of errors in the described reduced-order model.
Firstly, the model velocity field $\bm{u}'(\tv{x},t)$ may deviate from the true mean velocity $\bm{u}(\tv{x},t)$.
Secondly, the particle velocities may deviate from the mean velocity field.
One quantity that captures the level of fluctuations is the so-called \emph{granular temperature}, $T(\tv{x},t) = \left< \left[ \bm{v}_a - \bm{u}(\tv{x},t) \right]^2 \right>$, where  $\left< \hdots \right>$ denote averaging over particles in a small volume, centred at $\bm{x}$.
For the mean squared deviation of the particle velocities from the model velocity field we observe
\begin{equation}
	\left< \left\lVert \bm{v}_a - \bm{u}' \right\rVert^2 \right> \leq
	\left< \left\lVert \bm{v}_a - \bm{u} \right\rVert^2 \right> +
	\left< \left\lVert \bm{u} - \bm{u}' \right\rVert^2 \right>.
\end{equation}
We therefore introduce the \emph{granular temperature error}
\begin{equation}\label{eq:granular_temperature_error}
  \mathcal{E}_T(t) \equiv \left[ \frac{1}{V_B}\int_{V_B} w(\bm{x})\frac{\left< \left\lVert \bm{v}_a - \bm{u} \right\rVert^2 \right>}{v^2_{0}} \mathrm{d}\rule{-0.05em}{1.9ex}^3\bm{x} \right]^{1/2}
\end{equation}
and the \emph{velocity error} associated with the model reduction
\begin{equation}\label{eq:model_reduction_error}
	\mathcal{E}_v(t) \equiv \left[ \frac{1}{V_B}\int_{V_B} w(\bm{x})\frac{\left< \left\lVert \bm{u} - \bm{u}' \right\rVert^2 \right>}{v^2_{0}}  \mathrm{d}\rule{-0.05em}{1.9ex}^3\bm{x} \right]^{1/2}
\end{equation}
where the integrals are over the volume $V_B$ enclosing the reduced subsystem $B$ and $v_0$ is a characteristic velocity for the system that the model should be able to resolve.
The errors can be computed with no weight,  $w(\bm{x}) = 1$, or weighted by the local mass density, $w(\bm{x}) = \rho(\bm{x}) / \rho_0$, relative to a nominal bulk density $\rho_0$ to suppress harmless errors in dilute region.
If a surface height function $z = h(x,y)$ is tracked during a simulation, it can be interesting to analyse the \emph{surface height error}
\begin{equation}\label{eq:surface_error}
	\mathcal{E}_h(t) \equiv \left[ \frac{1}{V}\int_{A} \left[ h(x,y) - h'(x,y) \right]^2 \mathrm{d}\rule{-0.05em}{1.9ex}A \right]^{1/2}
\end{equation}
where $h'(x,y)$ is the surface height function using the reduced model, $A$ is the projected area of the system in the $xy$ plane and $V$ is the volume enclosed by $h(x,y)$ and some reference surface $h_0(x,y)$. 

\subsection{Extension to multibody systems}
Consider the presence also of a rigid multibody system $C$ with position $\bm{x}_C$, velocity $\bm{v}_C$ and mass $\bm{M}_C$.
The multibody system has articulation joints and actuators that are represented by a constraint vector $\bm{g}_C(\bm{x}_C,\bm{v}_C,t) = 0$ with Jacobian $\bm{G}_C = \partial \bm{g}_C/\partial \bm{x}_C$.
The forces on the multibody system are the constraint force $\bm{G}_C \bm{\lambda}_C$, external force $\bm{f}_C$ and contact forces $\bm{f}_{CA}$ and $\bm{f}_{CB}$ from the resolved system $A$ and the reduced system $B$.
The extended system has
the following equations of motion
\begin{align}
	\begin{bmatrix}
		\dot{\bm{x}}_A \\
        \dot{\bm{x}}_B \\
        \dot{\bm{x}}_C
	\end{bmatrix}
	& =
	\begin{bmatrix}
		\bm{v}_A \\
		\bm{u}(\bm{x}_B,t)\\
		\bm{v}_C
	\end{bmatrix}
	\label{eq:x_evolution} \\
	\begin{bmatrix}
	  \bm{M}_A \dot{\bm{v}}_A \\
	  \bm{M}_C \dot{\bm{v}}_C
	\end{bmatrix}
	& =
	\begin{bmatrix}
		\bm{f}_A + \bm{f}_{AB} + \bm{f}_{AC}\\
		\bm{f}_{C} + \bm{f}_{CA} + \bm{f}_{CB} + \bm{G}_C \bm{\lambda}_C
	\end{bmatrix}
	\\
        0 & = \bm{g}_C(\bm{x}_C,\bm{v}_C,t).
\end{align}
The multibody system has some velocity $\bm{v}_{BC}$ at the interface between system $B$ and $C$, which is a contributing cause of the velocity field $\bm{u}(\tv{x},t)$ in $B$.
The contact force $\bm{f}_{CB}$ on system $C$ from $B$ is either computed from a contact model or as an additional output of the reduced-order model.

\subsection{Adaptively reduced DEM}
\label{sec:adaptive}
The purpose of the model reduction is fast simulation with sufficient accuracy.
Fast simulation is achieved by minimizing the number of dynamic particles, $N^A_\text{p}$, that are simulated with high computational intensity in system $A$.
High accuracy requires that system $B$ does not include large domains with high granular temperature, where the individual particle motion deviate substantially from the mean flow.
The solution is to adaptively control which regions and what particles are simulated with the reduced DEM and resolved DEM, keeping the reduction factor $R$ and the effect of the granular temperature error $\mathcal{E}_T(t)$ at minimum.
This is carried out as follows, with reference to the illustration in Fig.~\ref{fig:system_illustration}.

The reduced model velocity field, $\bm{u}'(\tv{x},t)$, is assumed to be known.
Each particle $a$ has a speed $\Delta v_a = \left\lVert \bm{v}_a - \bm{u}' \right\rVert$ relative to the velocity field.
Particles are kept dynamic and part of system $A$ as long as their relative speed exceeds a threshold value $\Delta v_a > \varepsilon v_\text{0}$ for some error tolerance $\varepsilon$, which is application specific.
Particles with relative speed below the threshold value, $\Delta v_a \leq \varepsilon v_0$, are simulated with the reduced model in system $B$.
Now, contacts in granular media are strongly dissipative.
Consequently, particles in $A$ that repeatedly collide with particles in $B$ will have a velocity that quickly approach the velocity field and become part of system $B$.

It is easy to conceive extensions to this basic scheme.
Particles in $B$ that are impacted can be made dynamic and part of system $A$.
If the received impulse is large enough, and the particle lack contact support, it will no longer co-move with the velocity field and remain dynamic.
Otherwise, it will merge back.
If it is possible to predict regions of high granular temperature, the particles there can be kept dynamic.
This is useful at outlets, belt conveyor endpoints or when releasing material from a digging tool, where the flow transitions quickly from solid or dense liquid phase into gaseous phase and free fall.
If the velocity field has positive divergence, or if the mass density decrease well below a nominal bulk density, that is a clear indication that the structural rigidity is lost. 
Particles in such regions are subject to gravitational acceleration, which will eventually cause energetic impacts and rapid increase in the granular temperature.

\section{A velocity field predictor}
\label{sec:predictor}

Reduced-order DEM simulation, as outlined in Sec.~\ref{sec:RODEM}, rely on a lower-order model for predicting the velocity field in the granular media.
In this section we present a regression model for discrete velocity field prediction and the techniques we use for sampling training and test data from resolved DEM simulations.

\subsection{Coarse-graining}
Coarse-graining is a technique for sampling and averaging particle states to obtain macroscopic fields.
The field values at any coordinate $\tv{x}$ is a weighted average of a discrete set of particle and contact variables in a neighbourhood controlled by a coarse-graining function $\phi(\tv{x},R)$ with smoothing length $R$.
For sampling velocity fields when running the reduced-order model, we choose a Heaviside coarse-graining function.  For particles near the coarse-graining boundary, the mass is weighted by the volumetric overlap approximated using a bounding box.
In offline field analysis we use a Gaussian function $\phi(\tv{x},R) = (\sqrt{2\pi}R)^{-3} \exp (-|\tv{x}|^2/2R^2)$. 
Here we apply a cut-off at $|\tv{x}| = 3R$ for practical reasons, which cause a truncation error of $0.01$.  The mass density field is computed as $\rho(\tv{x},t) = \sum_a m^a \phi(\tv{x} - \bm{x}^a(t),R)$.  The velocity field is obtained by $\bm{u}(\tv{x},t) = \bm{p}(\tv{x},t)/\rho(\tv{x},t)$, where the momentum density field is first computed as $\bm{p}(\tv{x},t) = \sum_a m^a \bm{v}^a \phi(\tv{x} - \bm{x}^a(t),R)$.

\subsection{Discretization}
We use a regular grid with $N_\text{v}$ cells, voxels, with side lengths $\bm{L}= (L_x, L_y, L_z)$, where we denote the shortest of these $L_{\text{min}}$.
Each voxel, indexed $\bm{i} = (i,j,k)$, has a centre point coordinate $\tv{x}_{\bm{i}}$.
The mass density and velocity fields are represented in discrete form by their values in the voxel centres, $\rho_{\bm{i}} = \rho(\tv{x}_{\bm{i}})$ and $\bm{u}_{\bm{i}} = \bm{u}(\tv{x}_{\bm{i}})$.
The grid size is limited by the particle diameters $d$ according to $L_{\text{min}} > d_{\text{max}}$, where $d_{\text{max}}$ is the largest particle.
The smoothing length for coarse-graining is equal (Heaviside) or somewhat larger (Gaussian) than the size of the voxels.

\subsection{Sampling}
Data is sampled from resolved DEM simulations.
The instantaneous velocity field at a time $t$ is stored in a vector $\bm{U}(t) = [\bm{u}_{\bm{i}}(t)] \in \mathbb{R}^{3 N_\text{v}}$, the mass density field in a vector $\bm{P}(t) = [\rho_{\bm{i}}(t)] \in \mathbb{R}^{N_\text{v}}$, and the control signal that drive the system, in a vector $\bm{J}(t) = [j_i (t)] \in \mathbb{R}^{N_\text{j}}$.
These time instances are referred to as \emph{snapshots}.
A data \emph{sample} is a time-average of $N_\tau$ snapshots between time $t_{n}$ and $t_{n + N_\tau - 1}$, i.e.,
\begin{equation}
	\bm{U}_n = \frac{1}{N_\tau}\sum_{k = 0}^{N_\tau - 1} \bm{U}(t_{n + k}).
\end{equation}
Any reaction force on a selected body or contact surface, $b$, may be sampled as well. We denote this $\bm{F}(t) = [\bm{f}_b (t)] \in \mathbb{R}^{N_\text{f}}$, where $N_\text{f}$ is the total number of sampled force components.

\subsection{Regression model}
We are searching for a model that predict the
discrete velocity field $\bm{U}\in \mathbb{R}^{3 N_\text{v}}$, and possibly also the reaction force $\bm{F}\in \mathbb{R}^{N_\text{f}}$, from a given mass density field $\bm{P} \in \mathbb{R}^{N_\text{v}}$ and control signal $\bm{J} \in \mathbb{R}^{N_\text{j}}$.
This is approached as a regression problem, $\bm{y} = \bm{\phi}(\bm{x})$, with predictor variable $\bm{x} = [\bm{P}, \bm{J}] \in \mathbb{R}^{N_\text{v} + N_\text{j}}$ and response variable $\bm{y} = \bm{U} \in \mathbb{R}^{3N_\text{v}}$ or $\bm{y} = \bm{F} \in \mathbb{R}^{N_\text{f}}$ for the velocity and force prediction, respectively.  The natural start is to first consider a linear regression model
\begin{equation}
	\bm{\phi}(\bm{x}) = \bm{\beta}_0 + \bm{\beta}_1 \bm{x} + \bm{\varepsilon}
\end{equation}
with model parameters $\bm{\beta}_0 \in \mathbb{R}^{3N_\text{v}}$ and $\bm{\beta}_1 \in \mathbb{R}^{3N_\text{v} \times (N_\text{v} + N_j)}$ and error term $\bm{\varepsilon}$, for the case of the velocity response variable.
There is, however, good reason to believe that a purely linear model cannot capture the behaviour and our numerical experiments also confirmed this.
The velocity field and reaction force is expected to depend nonlinearly on the mass and the control signal. 
We assume that the velocity is linear to the control signal. Furthermore, we assume that the flow depends on the presence of material (voxel occupancy) rather than on the precise mass density. 
This lead to the following ansatz
\begin{equation}\label{eq:model_ansatz}
	\bm{\phi}(\bm{x}) = \bm{\beta}_0 + \bm{\beta}_1 \texttt{vec}[\hm{H}(\bm{P}) \bm{J}^T] + \bm{\varepsilon}
\end{equation}
where $\bm{H}:\mathbb{R}^{N_\text{v}} \to \mathbb{B}^{N_\text{v}}$ is the Heaviside function, component-wise returning an occupancy value $0$ or $1$ depending on whether the mass density in the voxel is nonvanishing.
The vectorization operator $\texttt{vec}(\ )$ produce a regression variable $\bm{x} \in \mathbb{R}^{N_\text{v} N_j}$ out of the matrix $\hm{H}(P) \bm{J}$ with dimension $N_\text{v} \times N_j$.
The model parameters is $\bm{\beta}_1 \in \mathbb{R}^{3N_\text{v} \times N_\text{v} N_j}$.
We make the same ansatz for the force response variable.

It can be expected that this model suffers from \emph{multicollinearity}, i.e., 
there might be predictor variables that are strongly correlated in the measured data.
One way to handle this is to apply \emph{Ridge regression} which adds a penalty term $\lambda \left\lVert \bm{\beta} \right\rVert^2_2$ to the regression loss function, where $\bm{\beta}_0$ and $\bm{\beta}_1$ have been combined in $\bm{\beta}$ as is customary. The regression loss function becomes
\begin{equation}\label{eq:loss_function}
	\mathcal{L} = \left\lVert \bm{y} - \bm{\beta} \bm{x}\right\rVert^2_2 + \lambda \left\lVert \bm{\beta} \right\rVert^2_2
\end{equation}
with penalty parameter $\lambda$, that is a hyperparameter to be calibrated.
Another way to treat the multicollinearity would be to apply \emph{principal component regression}.
In this case one performs PCA on the predictor variables and omit the low-order principal components.
Ridge regression accomplish the same effect, but without dimensional reduction.

\section{Numerical experiments}
\label{sec:numerical_experiments}
\begin{figure*} 		
    \centering
    \includegraphics[width=1.0\textwidth]{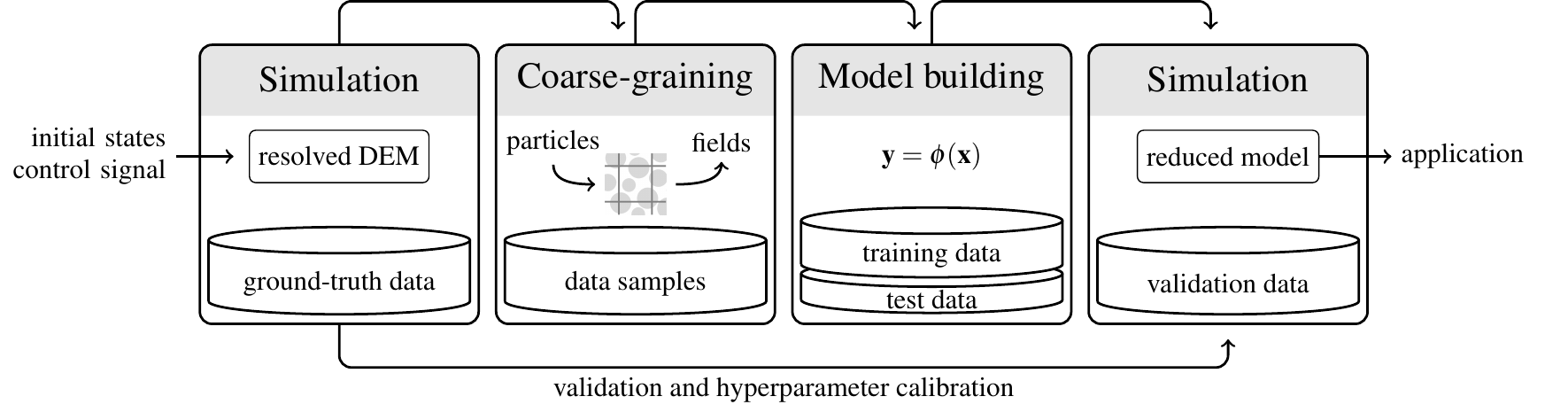}
    \caption{The different steps of developing a reduced-order model.}
    \label{fig:flow_chart}
\end{figure*}
To test the model order reduction technique, numerical experiments are performed on two systems.
We use a nonsmooth discrete element method as described in \cite{servin:2014:esn} using the software AGX Dynamics \cite{AGX}.
The reduced-order model is implemented in Python using NumPy and the model training is performed using scikit-learn \cite{scikit-learn}.
The experiment steps, summarized in Fig.~\ref{fig:flow_chart}, are as follows.
First, numerous ground-truth simulations are run.
System state snapshots are recorded, and coarse-grained data samples are produced.
Each data sample with index $n$ holds a discrete representation of the mass density field $\bm{P}_n$, velocity field $\bm{U}_n$, forces $\bm{F}_n$ and control signal $\bm{J}_n$.
The data samples are split in training data and test data by the ratio 80/20.
The model parameters, $\bm{\beta}$, that minimize the loss function, Eq.~(\ref{eq:loss_function}), are computed using \texttt{sklearn.linear\_model.Ridge}.
The regularization that give the best trade-off between training and test error is chosen manually.
The trained models are exported and used in DEM simulations for model order reduction.
Validation is made by running fully resolved ground-truth simulations and recording validation snapshots.  These are compared to snapshots recorded from reduced-order simulations starting from the same initial state and running with identical control signals.

\subsection{Pile with a discharge flow}
\label{sec:pile}
\begin{figure}
    \centering
    \includegraphics[height=0.6\columnwidth]{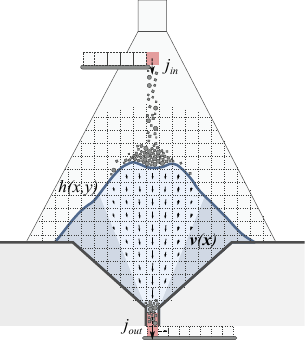}
    \includegraphics[trim={5mm 20mm 15mm 0},clip,height=0.53\columnwidth]{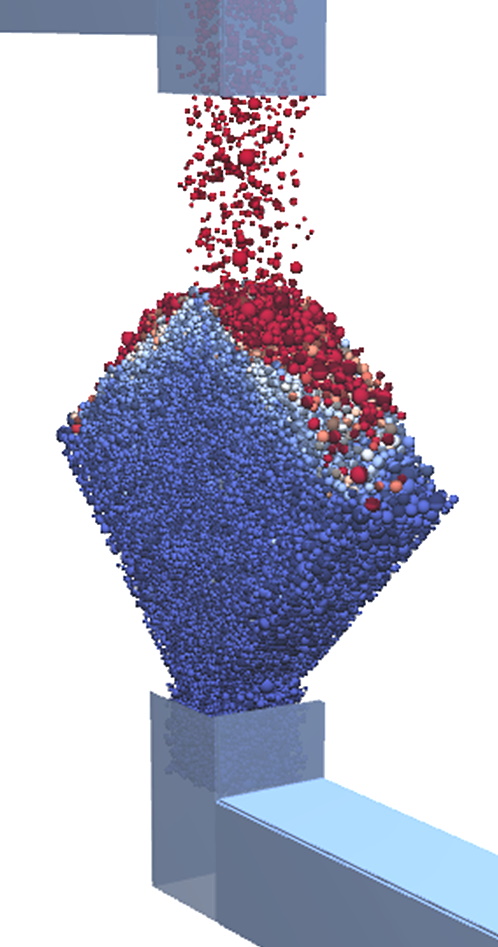}
    \caption{A pile with a discharge flow.}
    \label{fig:pile_system}
\end{figure}

A quasi 2D pile is confined by inclined sidewalls and vertical rear and front walls, as shown in Fig.~\ref{fig:pile_system}.
This represents a thin slice of a 3D system.
The inflow of material is controlled by an emitter above the pile, feeding material at variable flow rate.
There is a $1.9$~m wide outlet where the sidewalls meet. This is also the distance between the front and rear walls.
The sidewalls are inclined $48^{\circ}$ up to a plateau, where the distance between the sidewalls is 19.9~m.
The discharge flow at the outlet is controlled with a control signal $j(t)$.
Particles become kinematic at the outlet, moving with a velocity $\bm{v}_\text{out} = [0,0,-j(t)]$.

The particles are spherical with diameter 0.1~m, 0.16~m, 0.22~m, and 0.3~m, distributed by the mass ratio of 0.3, 0.5, 0.15, and 0.05, respectively, relative to the total mass. 
The specific mass density is $2500$ kg/m$^3$, elasticity $10^8$ Pa, coefficient of restitution $0.0$, friction coefficient $0.5$ and rolling resistance coefficient $0.2$.
The sidewalls share the same contact parameters as the particles.
Frictionless boundary condition is applied on the vertical (front/rear) walls.
The simulations are run with $200$ projected Gauss-Seidel solver iterations and  $0.017$~s time-step, set to avoid particles from tunneling through each other under the (time-implicit) numerical integration.
Grid dimensions are $24 \times 1.9 \times 15$~m with $40\times 3 \times 25$ voxels.
The predicted velocity field is a mean field, trained on coarse-grained data that involve both spatial and temporal averaging.
Consequently, the particles in the reduced domain may be integrated with a different time-step than used in resolved DEM simulation.
The Courant-Friedrichs–Lewy condition imply a time-step around 1~s or smaller for the given voxel size and flow rates.
We use $0.17$~s time-step for integrating the reduced-order model, which is a factor 3.5 below the CLF condition.
For the adaptive reduced DEM, velocities are predicted with a timestep of $0.17$~s, but particle positions are integrated at the same frequency as the resolved DEM.

Depending on the confinement geometry and material parameters, the granular media in a pile or silo is discharged either through funnel flow or mass flow.
In funnel flow, the material divides into stagnant zones with no motion, and flow zones with shear flow stretching from the outlet to the surface of the pile.
In mass flow there are no stagnant zones and all particles are in motion during discharge.
The pile in Fig.~\ref{fig:pile_system} exhibits funnel flow during discharge.
With a steep enough angle on the foundation, or small enough friction, this would instead have been mass flow.
For modelling the velocity field, we assume that the bulk flow is quasi-stationary and depend only on the current outflow control signal $j(t)$ and on the mass distribution $\rho(\tv{x},t)$.
Since the model converts the mass density into binary occupancy, the model can generally take the surface height function, $h(x,y)$, as input and directly compute the occupancy underneath it.
This is useful when running the model coupled to a real system instrumented with range sensors.

\begin{figure}
  \centering
  \includegraphics[width=0.45\textwidth]{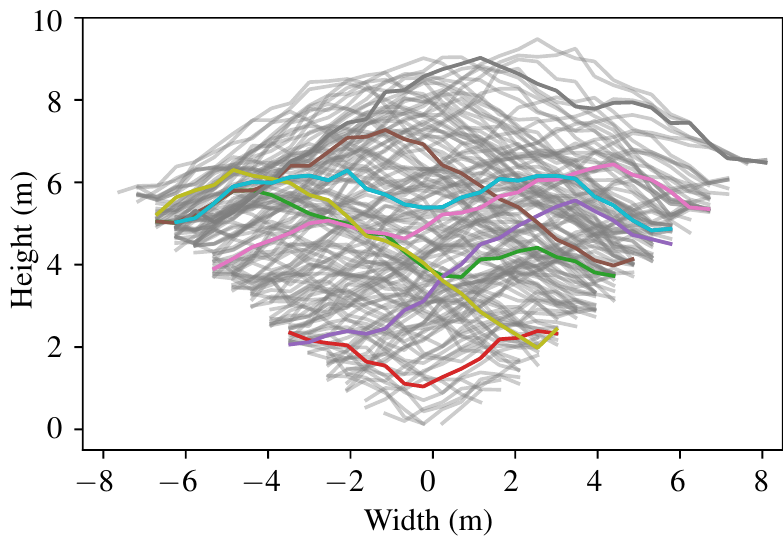}
    \caption{Distribution of surface profiles for the 150 initial states with highlighting of some randomly selected samples, which exemplifies the variation.
    }
  \label{fig:surface-profile-pile}
\end{figure}

It is important that the training data cover the system state space, that is spanned by the vector $[\bm{v}(\tv{x}),\rho(\tv{x}),j]$.
We generate data samples from 2500 full-resolution simulations, with varying outlet velocity, starting from 150 different initial states.
The outlet velocity is varied in the range 0~m/s to 0.5~m/s.
An initial state is a certain material distribution, created by different combinations of flow rates at the outlet and inlet.
Also, the position of the inlet is varied.
The surface profiles for the 150 initial states can be seen in Fig.~\ref{fig:surface-profile-pile}.

During each discharge simulation, the outlet velocity is kept constant at values between 0 and 0.5~m/s and there is no inflow.
The duration of each simulation is 30~s.  
Data samples with 1~s time average are produced from recorded snapshots. 
To avoid sampling any transient flow after engaging the outlet, the initial part of simulation is discarded.
The discharge simulations result in $\lesssim 50,000$ data samples, constructed with latin hypercube sampling uniformly from the 150 initial states and 2500 outlet velocities.
\begin{figure}
  \centering
  \includegraphics[width=0.45\textwidth]{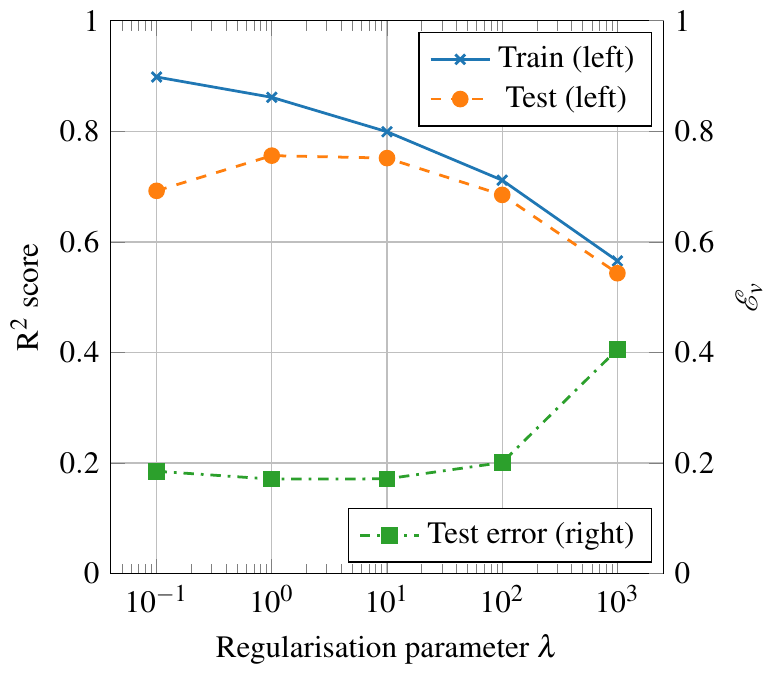}
  \caption{The $R^2$ score as a function of the regularization parameter $\lambda$ for the pile training- and test data (left), and the average model reduction velocity error for the test data (right).}
  \label{fig:model_parameters}
\end{figure}

A number of models for predicting the velocity field are generated with ridge regression parameter in the range $10^{-1}$ to $10^3$.
The performance of these models is evaluated in multiple ways.
i) The prediction score on the training and test data are compared to see how well the models can generalize to unseen states.
ii) The ability to predict velocity fields is examined by comparing coarse-grained resolved DEM fields to corresponding predicted fields, in a complete discharge of a chosen pile state.
iii) The same pile discharge is used to compare when each particle pass the outlet (exit time) in reduced-order DEM c.f. resolved DEM.
iv) The adaptive reduced-order DEM technique is evaluated by comparison to resolved DEM during more extensive filling and discharging with identical inflow/outflow signals.

The prediction score on the training and test data can be seen in Fig.~\ref{fig:model_parameters}.
The best performance on the unseen test data occurs for regularization parameter values between 1 and 10.
We chose $\lambda=10$ as our preferred model and will focus on the performance of this.

Fig.~\figref[a]{fig:pile_discharge_field_comparison} shows snapshots from a ground truth resolved DEM simulation of the discharge of a pile state with constant outflow velocity 0.5~m/s.
The columns are snapshots from 10, 20, 30 and 40 seconds into the simulation, with the fields time-averaged over 1~s.
In the third and fourth row the velocity field from the ground truth simulation and the model prediction ($\lambda=10$) are shown together with the mass density.
In the fifth row the difference between the ground truth and predicted velocity field is shown, and we observe a good agreement.
Fig.~\figref[b]{fig:pile_discharge_field_comparison} shows the model reduction velocity error, defined in Eq.~(\ref{eq:model_reduction_error}), as a function of time for the duration of the pile discharge.
This evaluates the performance of the velocity prediction for all the produced models.
We can see that the model reduction velocity error is around $10\%$ for the $\lambda=10$ model but increases to $20\%$ towards the end when the amount of remaining material is small.
In the second row we observe that the granular temperature error is elevated especially near the outlet, indicating an irregular flow there.
The time-evolution of the granular temperature error is also included in Fig.~\figref[b]{fig:pile_discharge_field_comparison}.
It varies mostly between $0.5$ and $1$. 

\begin{figure}
  \centering
  \includegraphics[width=0.5\textwidth]{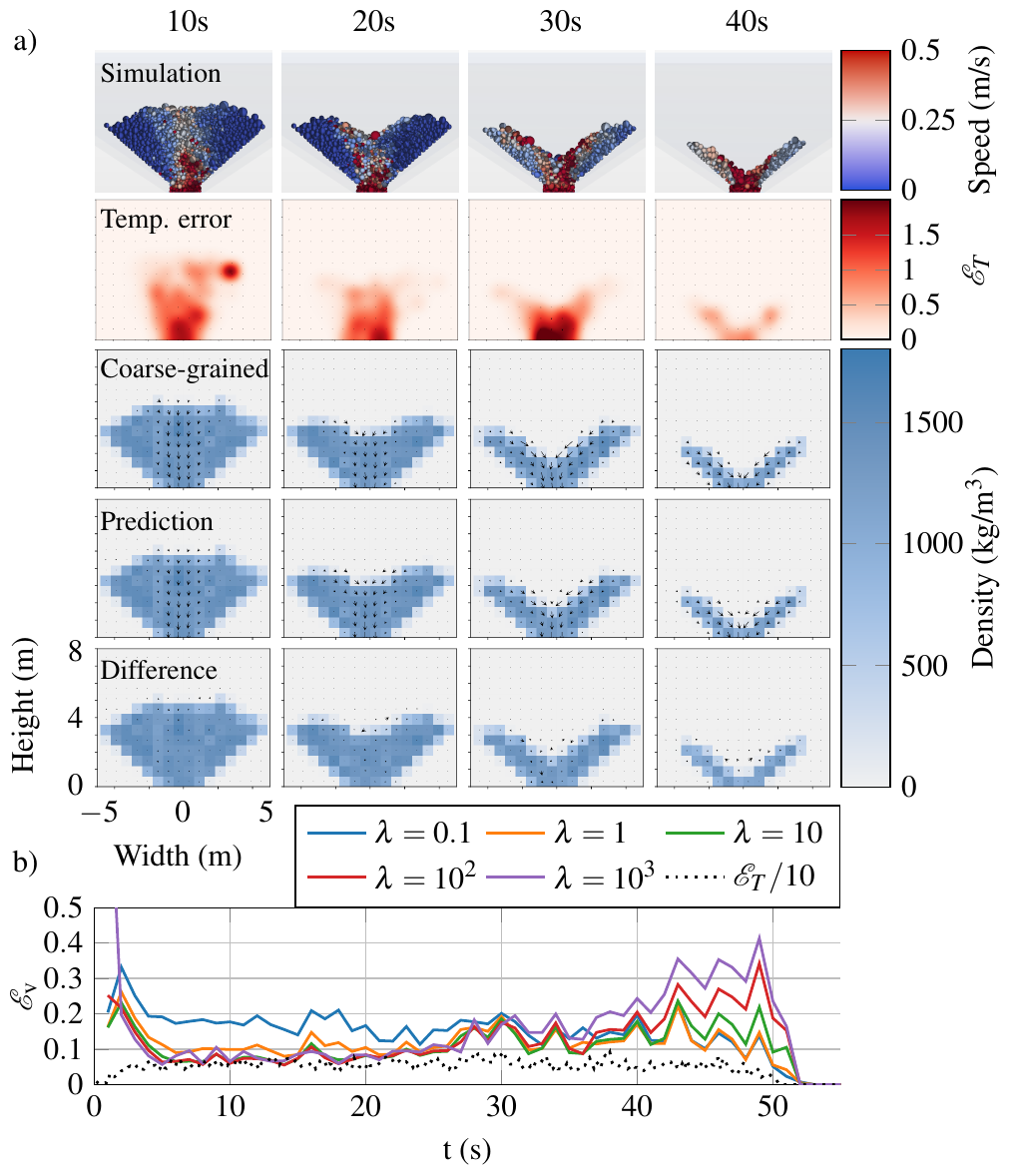}
  \caption{a) Sample results, comparing ground truth velocity fields from DEM simulations with predicted velocity fields from the reduced-order model. b) The model reduction velocity error.}
    \label{fig:pile_discharge_field_comparison}
\end{figure}
\begin{figure}
    \centering
    \includegraphics[width=0.5\textwidth]{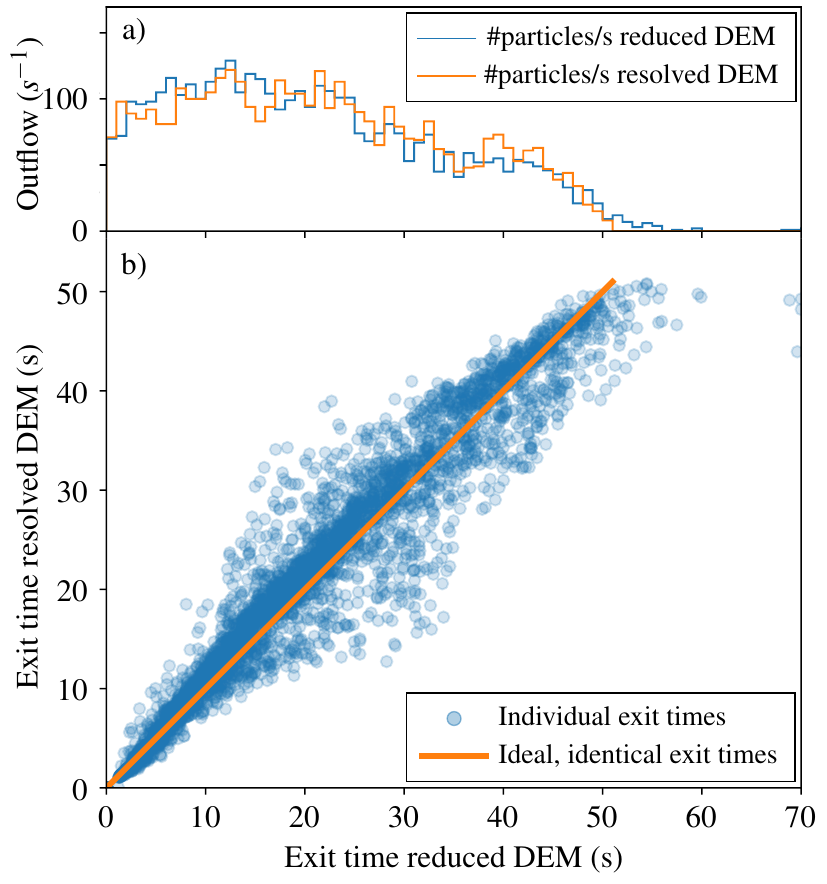}
    \caption{Comparison of residence times from pile discharge using fully resolved DEM simulations and the reduced-order model. a) Number of discharged particles per unit time for the two cases, i.e. the marginal distributions of b) individual discharge times for each particle in the two cases.
}
    \label{fig:residence_time_error}
\end{figure}

The ability to predict the velocity field is necessary for reduced-order DEM simulation, but not sufficient.
When the system is time-integrated using the reduced-order model the errors may drift or cause instability.
Therefore, we examine also the performance of the model when used to propagate the system forwards in time during a pile discharge.
Starting from the same state as in Fig.~\ref{fig:pile_discharge_field_comparison} and running with the same control signal, particle positions are integrated using linear interpolation of the velocity field to the particle positions.
We compare the particles exit time in the two cases, the ground truth resolved DEM simulation and the reduced-order DEM case.
The results can be seen in Fig.~\ref{fig:residence_time_error}.
The orange line indicates identical exit times for the two cases.
Most particles are concentrated on this line, with standard deviation $3.4$~s and mean absolute deviation $2.1$~s, which is a relative error around 10\%.
The distributions of number of outflow particles per time unit are also in fair agreement.  

Finally, we evaluate the performance of the adaptive reduced-order DEM technique.
Here, reduced DEM is used to simulate the motion below the surface of the pile, where we expect the media to be in the liquid or solid phase. 
Resolved DEM is used for the particles in free fall, impacting and flowing rapidly on the surface, i.e., in the gaseous phase.
The idea is to run one resolved DEM simulation and one adaptive reduced-order DEM simulation using identical control signals.
The tests start from an empty container, building up a pile with variable inlet and outlet signals.
The control signals can be seen in Fig.~\figref[b]{fig:pile_demo_comparison}, with the outflow velocity converted to an estimated mass outflow per unit time.
The outflow velocity is initially kept within the domain of the training data (0 to 0.5~m/s), but is in the end set well outside this domain, at 1~m/s.
\begin{figure}
    \centering
    \includegraphics[width=0.5\textwidth]{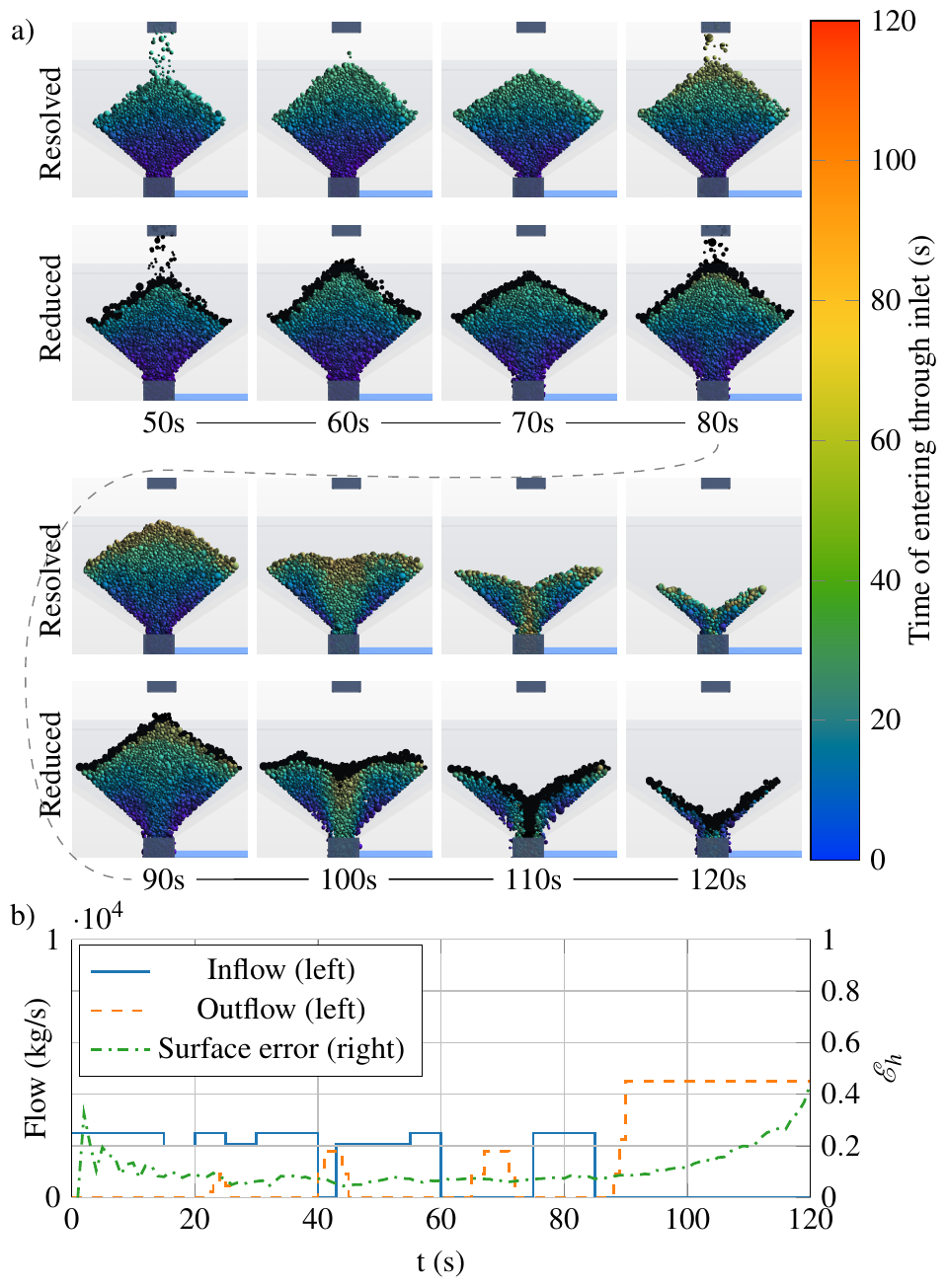}
    \caption{a) Comparing ground truth simulations and model reduction simulations with identical inflow-, and outflow signals.
      b) Rate of inflow and outflow, as well as the surface height error.
      Note that the reason particles can be seen passing through the outlet wall in the reduced DEM case is that the boundaries of the voxels are not matching the outlet geometry.
    }
    \label{fig:pile_demo_comparison}
\end{figure}
A comparison of the two cases can be seen in Fig.~\figref[a]{fig:pile_demo_comparison}, with snapshots at every 10 seconds after building up the pile.
The fully resolved DEM simulations (ground truth) can be seen above the corresponding adaptive reduced-order model case.
The particles are colour coded according to the time they enter through the inlet, with the dynamic particles in the adaptive reduced-order model case displayed in black.
We track the surface of the pile and let particles down to a depth of 0.5~m be dynamic.
The relative speed threshold value, for turning particles kinematic, is set to $0.01$~m/s.
One can see that the surface profiles are similar between the two cases for essentially all times.
The surface height error can be seen, as a function of time, in Fig.~\figref[b]{fig:pile_demo_comparison}.
It is well below $10\%$ during most of the simulations but increase towards the end when there is less remaining material.
Since it is a relative error, this increase has limited practical implication.

\begin{figure}
    \centering
    \includegraphics[width=0.5\textwidth]{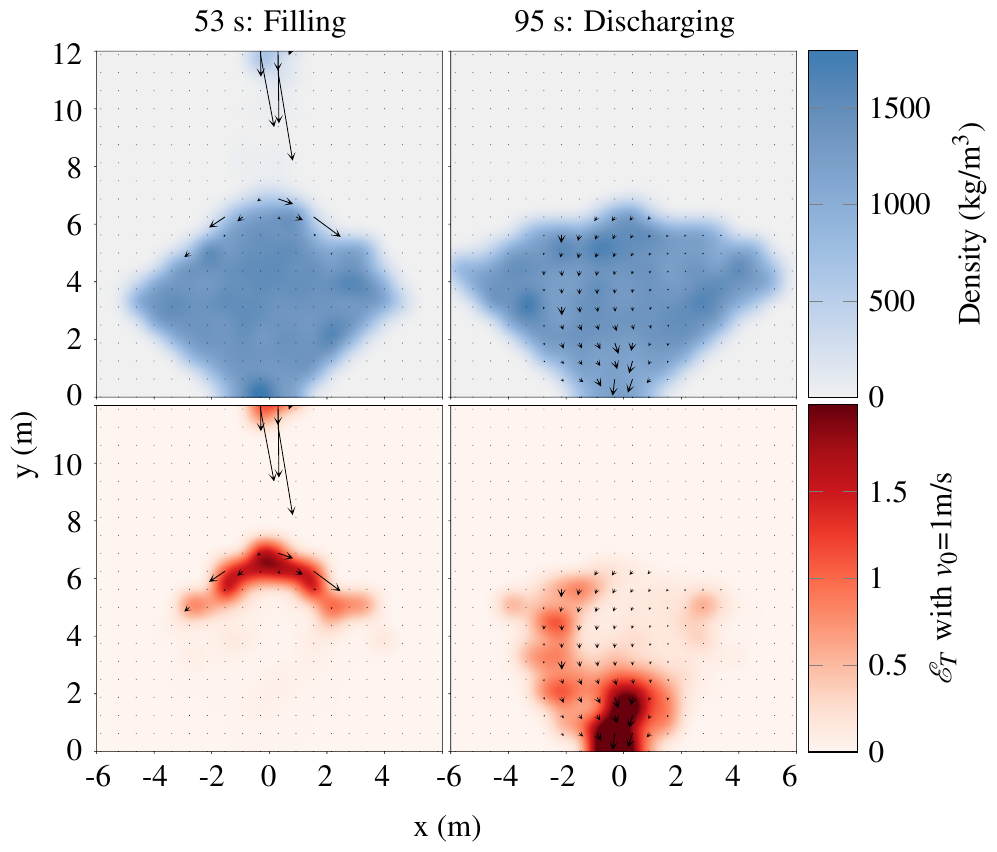}
    \caption{The temperature error (bottom row) from a full resolution simulation of a pile during filling (at 53~s, left column) and discharging (at 95~s, right column). 
      The velocity vector field and mass density field (top row) are included for reference.
    }
    \label{fig:granular_temperature_error_pile}
\end{figure}

It is interesting to study the granular temperature error, defined in Eq.~(\ref{eq:granular_temperature_error}), to gain insight in the deviation of particle motion simulated with the full resolution model and the reduced-order model.  Two snapshots, from time 53~s and 95~s, are presented in Fig.~\ref{fig:granular_temperature_error_pile}.
As expected, the granular temperature error is elevated on the surface of the pile when there is an incoming flow. 
This confirms our assumption that the material is in the gaseous phase, motivating the use of resolved DEM there.
During discharge, the granular temperature error is elevated around the outlet, not surprising given that the largest particle diameter is 1/6 of the outlet width. 
There are also signs of correlated velocity fluctuations on a longer length and timescale than that of individual particle rearrangements.
This limits the possibility for the reduced models to achieve a low velocity error $\mathcal{E}_v$, even if the model accurately predicts the mean fields.
Since this occurs near the outlet, it has marginal effect on the material above.

Sample videos are available at \url{http://umit.cs.umu.se/ddgranular/}.

\subsection{Bulldozing blade}
\begin{figure}
    \centering
    \includegraphics[width=0.5\textwidth]{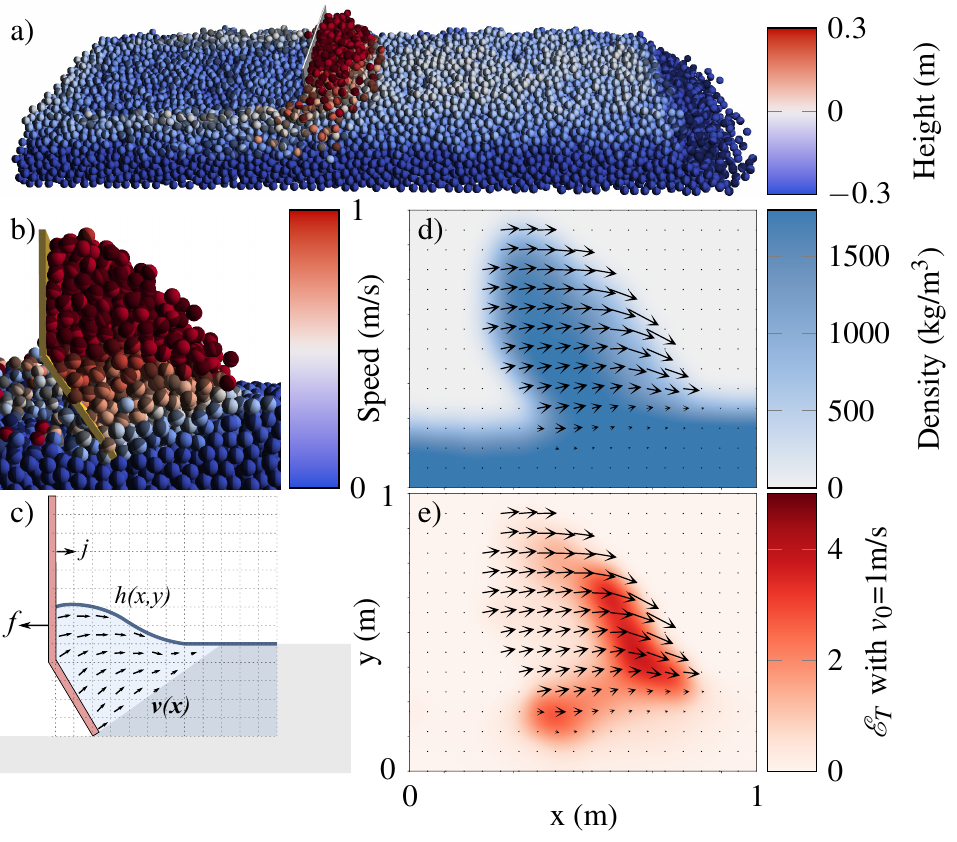}
    \caption{A blade pushing a granular bed simulated with fully resolved DEM.
A time instance is shown in 3D overview (a) and cross-section view (b).
The predictor model for the velocity field and reaction force from the control signal and mass distribution is illustrated in (c).
Also shown is the mass density (d) and granular temperature error (e), with the mean velocity vector field superimposed.}
    \label{fig:blade_system}
\end{figure}

As a second test we simulate a blade driven horizontally, cutting the surface of a granular bed like a bulldozer blade, see Fig.~\ref{fig:blade_system}.
The simulated flow is consistent with the theory of soil mechanics, that predict the formation of a wedge-shaped failure zone in front of the blade \cite{mckyes:1985:sct}.  
When pushed forward, the soil fail along a localized shear band that stretch from the cutting edge of the blade up to the free surface.
Outside the failure zone the material is at rest.  Inside the failure zone the material moves forward and upward, and may form a pile with a circulating flow.
The granular temperature is elevated at the front surface of the pile and at the cutting edge where impacts are frequent.
A model is trained to predict the velocity field and the reaction force on the blade from the horizontal velocity and the mass distribution in front of the blade.
The blade is $1.6$~m wide and is attached with a 6-degree-of-freedom constraint to a kinematic body having velocity $\bm{v} = [j (t), 0, v_z (t)] $.
The constraint force holding the blade relative to the kinematic body $\bm{f} = [f_x, f_y, f_z] $ is measured during the simulation.
The shape of the blade is that of two rectangular plates joined along their long edge at an angle of 35$^{\circ}$.
To avoid sampling of an unnecessarily large domain a coarse-graining grid is co-moving with the blade as in Fig.~\figref[c]{fig:blade_system}, with sampled quantities in the world frame coordinates.
The dimensions of the grid is $1.0 \times 2.5 \times 1.0$~m with $8\times 5 \times 8$ voxels.
Data samples are time-averaged over 1/6~s, chosen to remove velocity fluctuations while still resolving the motion of the blade.
In this numerical experiment the ground truth simulations are run with a time-step of $0.005$~s and 250 projected Gauss-Seidel solver iterations. The particle contact parameters are identical to the pile experiment, but the particle diameter is set to $10$ cm.

During simulations using the adaptive model order reduction, the particles inside the co-moving grid are assigned the velocity predicted by the velocity field using linear interpolation between the voxel centres.
Outside the grid, the velocity field is assumed to be zero.
Particles exiting the co-moving grid become dynamic, simulated using resolved DEM, until they are at relative rest to the particle bed.

\begin{figure}
  \centering
  \includegraphics[width=0.45\textwidth]{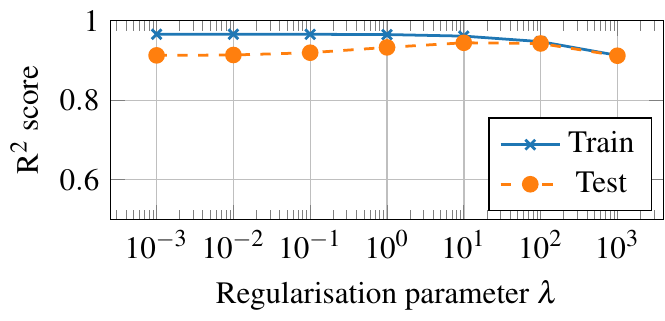}
  \caption{The $R^2$ score as a function of the regularization parameter $\lambda$ for the blade training- and test data.
}
  \label{fig:model_parameters_blade}
\end{figure}

To generate data samples, 250 simulations are run where the blade is pushed with different constant velocities, between 0 and 1.5~m/s, and with different cutting depth, ranging between 0 and 0.2~m.
The model is trained using exclusively horizontal motion.
As for the pile, a number of models are generated with different ridge regression parameter values, here ranging from $10^{-3}$ to $10^{3}$.
The prediction score on the training and test data can be seen in Fig.~\ref{fig:model_parameters_blade} and we again pick $\lambda=10$ as our preferred model, with the best generalization to the test data.

The models are evaluated with regard to i) the ability to predict velocity fields, ii) the ability to predict the force holding the blade, and iii) the ability to propagate the system forward in time.
This is all considered for a standardized blade trajectory, which can be seen in Fig.~\figref[a]{fig:blade_force}.
\begin{figure}
    \centering
    \hspace{5mm}
    \includegraphics[width=0.5\textwidth]{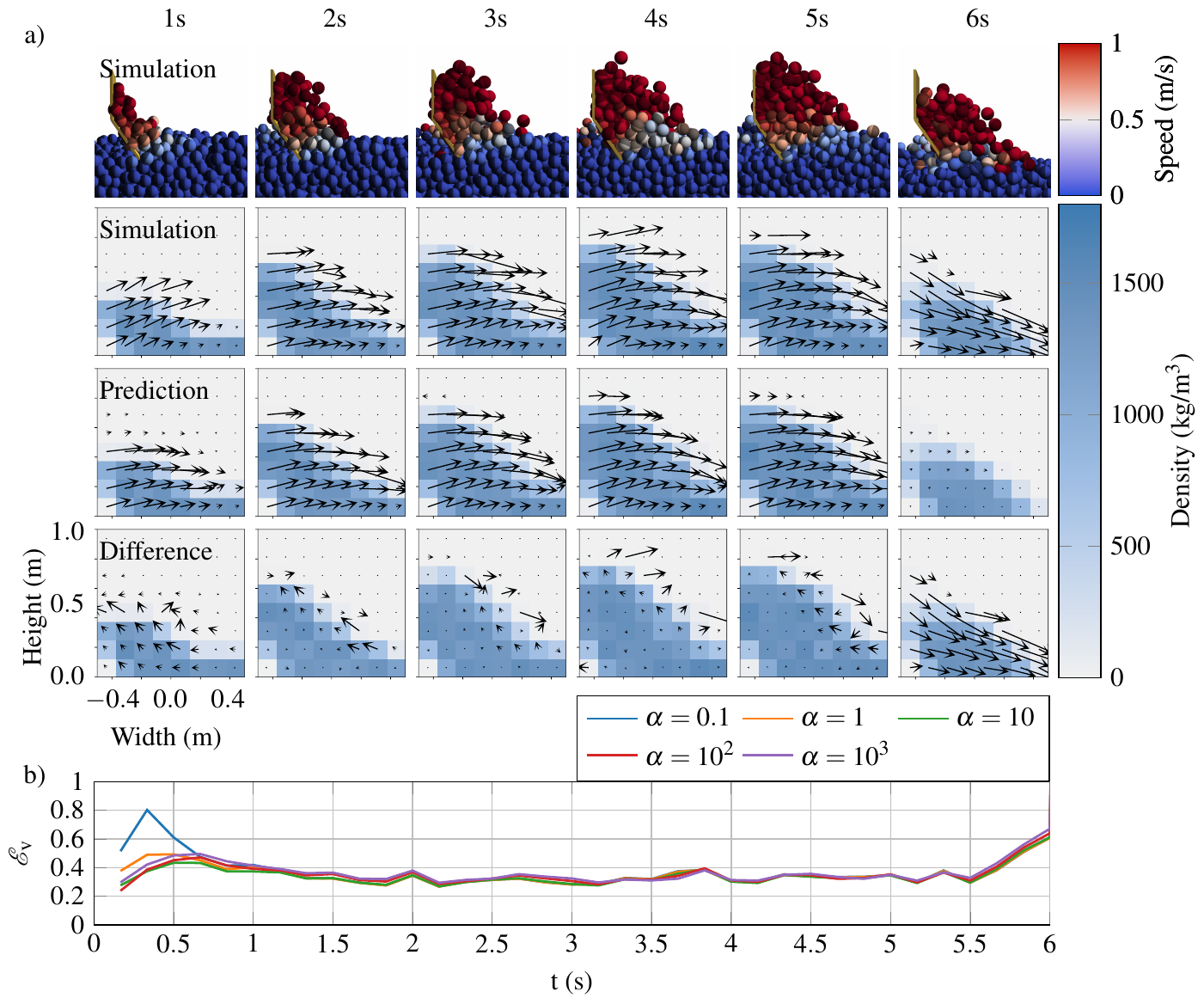}
    \caption{a) Sample results, comparing ground truth velocity fields with the velocity fields predicted by the reduced-order model.
b) The model reduction velocity error.
}
    \label{fig:blade_array}
\end{figure}

The ability to predict the velocity fields is evaluated by comparing coarse-grained ground truth velocity fields to that predicted from the corresponding mass densities, as seen in Fig.~\ref{fig:blade_array}.
The first row show states at 1~s intervals from a resolved DEM simulation with the blade cutting soil at 1~m/s and 0.15~m depth.
The second row show the corresponding mass density and velocity field obtained by coarse-graining.
The velocity field predicted by the trained model is shown in the third row, and the fourth row show the difference between the simulated and predicted velocity fields.
The predicted velocity fields are generally close to the measured ones inside the active zone but can differ considerably on the surface.
This is consistent with the observation, in Fig.~\figref[e]{fig:blade_system}, that the granular temperature is elevated there, indicating material in the gaseous phase.
This suggests using resolved DEM for the particles on the surface to minimize the reduction error, which however, was not employed in this test.
The evolution of the model reduction velocity error, $\mathcal{E}_\text{v}$, over time and for the different regularization parameters is shown in the bottom of the figure.
For the $\lambda = 10$ model the error is at $0.35$.
The errors are the largest around 1s and at the end, when the blade is raised.
This is not surprising as this was not included in the training data.

The capability of predicting the force required to push the blade through the particle bed is also examined.
A fully resolved DEM simulation is performed using the described trajectory to produce ground truth data of the blade force.
From the same simulation, mass density field and control signal are used as input to the model for predicting the blade force.
Both forces, time-averaged over 1/6~s, are plotted in Fig.~\figref[b]{fig:blade_force}. 
The path of the blade (top figure) is illustrated with 1~s intervals.
\begin{figure}
    \centering
    \includegraphics[width=0.45\textwidth]{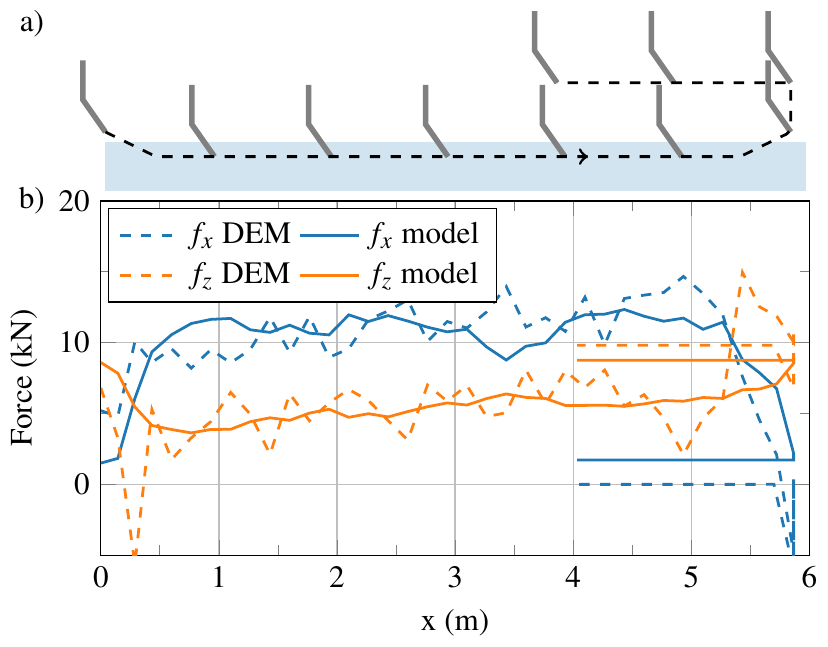}
    \caption{
      a) Motion of blade, along dashed black line, with blade profiles at 1~s interval in gray, and soil in blue.
      b) Sample data of the horizontal (blue) and vertical (orange) components of the force to push the blade, simulated using resolved DEM (dashed lines) and the reduced model (solid lines).
    }
    \label{fig:blade_force}
\end{figure}
The predicted force is overall in good agreement with the ground truth but has some problems when the blade is lowered into and raised from the bed, which was not represented in the training data.
Also, it was found that training data with the blade moving above the bed, without any mass in sampling region, was necessary for the model to predict the weight of the blade, i.e., the vertical force when the blade is reversed at the end of the bulldozing cycle. 

\begin{figure}
    \centering
    \includegraphics[width=0.45\textwidth]{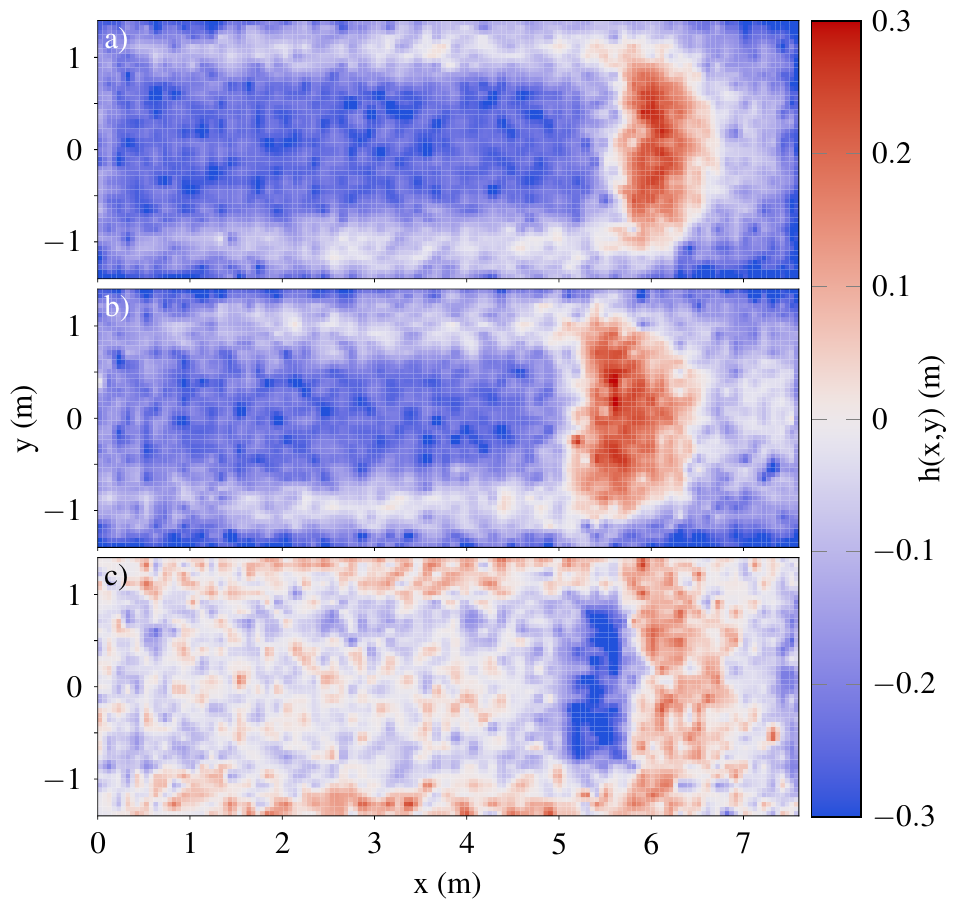}
    \caption{The surface of the material bed after completion of one bulldozing cycle (at 9 s) with ground truth simulation (top), granular reduced-order model (middle) and the difference between them (bottom).
}
    \label{fig:blade_heightprofile}
\end{figure}

Predicting the velocity field is a necessary but not sufficient functionality. 
It does not imply that the reduced-order model can propagate the system forward in time with similar accuracy. 
Velocity errors in the build up phase may lead to unphysical material distributions, in which case it is of little value that the model can predict stationary flows.
Also, there are regions of elevated granular temperature in Fig.~\figref[e]{fig:blade_system}, at the blade's cutting edge and the front surface of the pile.
Since the particle velocities deviate from the mean flow in those regions, the model should not be able to predict them accurately.
Therefore, we compare reduced-order DEM and resolved DEM simulations with identical blade trajectories, to evaluate the performance in propagating the system forwards in time.
The reduced-order DEM is simulated with the same blade trajectory as previously, depicted in Fig.~\figref[a]{fig:blade_force},
but with the larger time-step $0.017$~s.
Fig.~\ref{fig:blade_heightprofile} shows the two surfaces after a completed bulldozing cycle, and the difference between them.
The surfaces have slots and side windrows of similar depth, height and width.
In the reduced DEM simulation, the resulting pile is not pushed as far to the end of the slot as in the resolved DEM simulation.
This can be understood by that the particles have no inertia in the reduced-order model, and the predictor do not consider the vertical motion when the blade is lifted.
Instead, the material simply stays still when the blade is stopped and lifted, and subsequently fall down in a pile when exiting the voxel field.

The reduced-order model captures the force and the material displacement fairly well but the particle velocities only to partial extent.
It is also found that the model is sensitive to the training data.
The problem is that the flow during the build up phase is rather different from the stationary flow, and it turns out it is hard to find a balance in representing them both with the model.
Depending on which temporal parts of the training data are included, more or less emphasis can be put on these parts.
This imbalance typically results either in particles not building up properly in front of the blade, or particles not able to comove with the blade, leaking out through the back.
The velocity error shown in Fig.~\ref{fig:blade_array} at time 1~s illustrate the former.

Sample videos are available at \url{http://umit.cs.umu.se/ddgranular/}.

\subsection{Performance measurements}
In Table \ref{tab:runtimes_pile} we summarize the performance measurements from the numerical experiments.  
The computational time for the fully resolved and reduced DEM simulations, $t_\text{res}$ and $t_\text{red}$, are normalized by the real time duration of the experiments, $t_\text{real}$.
The speed-up factor is the ratio of the resolved computational time to the reduced one.
The reduction factor is the ratio of the average number of dynamic particles in the reduced and the resolved DEM simulations.
The accuracy is the errors subtracted from unity.

The numerical experiments are performed on a prototype implementation, combining Python and NumPy operations for the reduced-order DEM and AGX Dynamics for resolved DEM and multibody dynamics.
We expect there is room for optimization of the computational speed.
One opportunity that has not been utilized is that the number of iterations in the projected Gauss-Seidel solver can be decreased with the number of resolved particles (size of the contact network) \cite{servin:2014:esn}.
In the performed tests the number of solver iterations was kept constant at 200 and 250 in the pile and the blade experiments, respectively. 
If the reduced DEM simulations instead are run with 20 solver iterations, appropriate for an error tolerance of 10\% on these systems, the speed-up would roughly double from 10 to 20 (Experiment Pile Fig.~\ref{fig:pile_demo_comparison}) and from 50 to 100 (Experiment Blade Fig.~\ref{fig:blade_array}), respectively, and reach the real-time requirement $t_\text{red}/t_\text{real} < 1$.
Disabling the collision detection between particles in the reduced-order model is another optimization that has not been applied here.
The measurements of the computational time were made using a single thread on an Intel\textsuperscript{\textregistered} Core\textsuperscript{\texttrademark} i7-4770 CPU @ 3.40GHz.

\begin{table*}[htbp]

  \centering
  \small
  \begin{tabular}{l | l l l l l}
    Experiment & $t_\text{res}/t_\text{real}$ & $t_\text{red}/t_\text{real}$ & Speed-up & Reduction & Accuracy\\
    \hline
    Pile Fig.~\ref{fig:pile_discharge_field_comparison}& 271/70 & 13/70 & 21 & 0/1700 & 90\% \\
    Pile Fig.~\ref{fig:pile_demo_comparison}& 1496/120 & 149/120 & 10 & 440/3600 & 90\% \\
    Blade Fig.~\ref{fig:blade_array}& 630/8 & 11/8 & 57 & 250/16000 & 65\% \\
  \end{tabular}
  \caption{Performance number measured in the numerical experiments using resolved and reduced DEM simulations. }
  \label{tab:runtimes_pile}
\end{table*}

\section{Discussion}
The purpose of the model order reduction is to enable simulation at a higher speed or with increased number of particles while remaining within given computational bounds.
The price for the speed-up is reduced accuracy and the effort of the simulations that must be carried out in advance to produce training data.

Let us first consider what computational speed-up can be expected.
In the extreme case when the whole system can be represented with the reduced model, the main computational steps are: i) do inference on the regression model in Eq.~(\ref{eq:model_ansatz}) with given known parameters $\bm{\beta}$; ii) determine the particle velocities by interpolation and update their new position; iii) and, possibly, generate output for the purpose of analysis.
The computational complexity of doing inference is that of matrix-vector multiplication of size $\dim (\bm{\beta}) = 3N_\text{v} \times N_\text{v}N_\text{j}$,
which require $6 N_\text{v}^2N_\text{j}$ floating-point operations.
At the present time a powerful desktop CPU delivers about 100 gigaFLOPS and a high-end GPU up to 100 teraFLOPS.
It is thus conceivable to evaluate models of size $N_\text{v}=10^3$ (CPU) and $N_\text{v}=10^4$ (GPU) within one millisecond.
Hence, it should be possible to simulate fully reduced DEM systems at 60 Hz with up to $N_\text{p}=10^5$ (CPU) and $N_\text{p}=10^6$ (GPU) particles, assuming 10 particles per voxel. 
Also, the reduced-order DEM is limited by Courant-Friedrichs–Lewy (CFL) condition rather by the time-step used in simulation of the resolved DEM.
In the pile and blade experiments the CFL time-step limits are estimated to $1$~s and $0.1$~s, respectively.
That adds another factor 10 to 100 in speed or size of systems that may be simulated in real-time with the reduced model.

When the system has regions with granular temperature domains that require resolved DEM, this part of the simulation easily becomes the computational bottleneck.
The number of particles that can be simulated in real-time with resolved DEM is up to $10^4$, the precise number depending on particle size, velocity, contact parameters and numerical integration technique \cite{servin:2014:esn}.
With a reduction factor of 1/10, we expect that the pile system can be simulated in real-time with $N_\text{p} = 10^5$ particles.
Compared to this theoretical performance estimate, the prototype implementation is underperforming in the Pile test in Fig.~\ref{fig:pile_demo_comparison} by a factor of 10 and can potentially be optimised to reach a speed-up factor of 100. 

The accuracy in the pile experiments is about 10\%.
Whether this is acceptable depend highly on the application and if there are any alternatives for achieving the required simulation speed.
There are several ways the reduced model can be improved, besides providing it with more or better training data.
The velocity field in the real system has fluctuations and transients that cannot be captured by a model assuming quasi-stationary flow.
To achieve higher accuracy, the missing flow dynamics must be included either at the level of the velocity field predictor or at the particle level, like in the spot-model by Rycroft et al. \cite{Rycroft2006}.
The error due to elevated granular temperature in shear flow can possibly be reduced by adding diffusion terms in the calculation of new particle velocities from the mean field.
Models for diffusion in granular flows, as function of the local strain rate, can be found in the literature \cite{Utter2004} and can be calibrated using the resolved DEM simulation data.
It may also be a good idea to adjust the particle velocities or positions to resolve unphysical contact overlaps that result from the errors in the velocity predictor.
In the prototype implementation used for the numerical experiments, the regions of elevated granular temperature are detected manually.
Creating a model for predicting the local granular temperature, and how it can be expected to develop, would enable more automatic and accurate adaptive reduced-order DEM simulation.
The current paper focus on using a quasistationary velocity field for the reduced DEM.
It is a compelling idea, seemingly feasible, to extend this to viscoelastic deformations using 
the method of Zhong and Sun \cite{Zhong2018} for predicting the velocity field.

\begin{figure}
  \centering
  \includegraphics[width=0.45\textwidth]{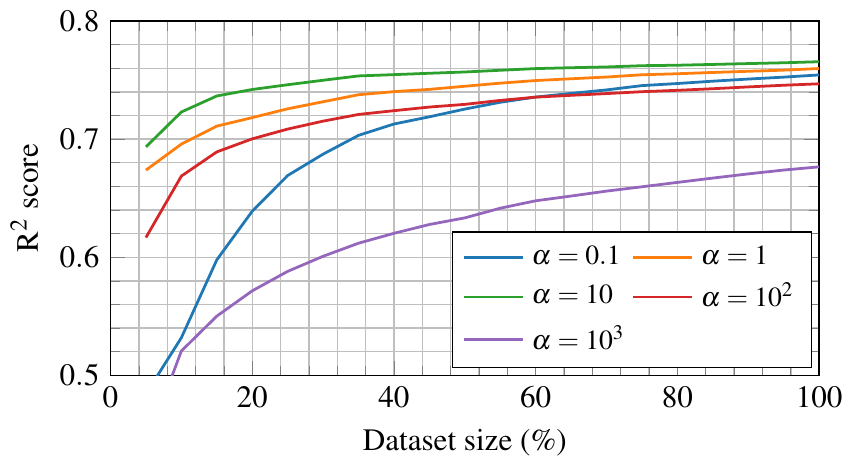}
  \caption{R$^2$ scores for the pile models on test data, trained on fractions of the training dataset.}
  \label{fig:training_data}
\end{figure}

Another important question is how much training data is needed to create the reduced-order model.
Fig.~\ref{fig:training_data} shows the prediction score on the test data for the pile models, as a function of being trained on increasing fractions of the training set of 50,000 data samples. 
The most significant increase in the test score occurred when using up to 50\% of the training set, with generally a small further increase when using all data.
However, the scaling depends on the regularization and none of the cases has really saturated, thus there could still be some improvements by increasing the amount of training data.

In the current paper we have not made any tests involving particle cohesion or adhesion. 
As long as there is a relationship between velocity field and control signal in the training data, we expect that reduced order models can be developed for this case as well.

\section{Conclusion}
We introduce a novel technique for model order reduction of DEM simulations of granular media.
Using many offline simulations, we train a regression model to predict the velocity field.
This model is then used to assign particle velocities, in place of the time-consuming process of collision detection and force computation.
An adaptive domain technique is used to apply the reduced-order model in regions with low granular temperature error where individual particle motion coincide well the mean flow, and use resolved DEM simulation in regions with more irregular particle motion.
This allows for minimizing the number of particles in the computationally intense process, while still simulating particle motion with realistic mean velocity.
The adaptive reduced-order model is applied in two test systems, a gravitational discharge pile and a bulldozer blade cutting and pushing through a particle bed.
We measure a computational speed-up of 10 to 20 and 90\% accuracy for the discharge pile.
We estimate that it is possible to reach real-time performance with $10^5$ particles at 60 Hz and a reduction factor $1/10$, corresponding to a speed-up of 100.
For the bulldozing blade the speed-up reaches 60 but with an accuracy of around 65\%.
Plans for future research includes extending the predictor model beyond plain regression.

\subsection*{Acknowledgements}
The project was funded in part by eSSENCE and VINNOVA (grant id 2019-04832) and has been supported by Algoryx Simulation AB.
The simulations were performed on resources provided by the Swedish National Infrastructure for Computing (SNIC dnr 2019/3-168) at High Performance Computing Center North (HPC2N).

\subsection*{Conflicts of interest}
Martin Servin is co-founder of and senior scientist at Algoryx Simulation AB.

\bibliographystyle{abbrv}

\end{document}